P9.7    RECENT DEVELOPMENTS IN MICROBURST NOWCASTING USING GOES

Kenneth L. Pryor
Center for Satellite Applications and Research (NOAA/NESDIS)
Camp Springs, MD

## 1. INTRODUCTION

Recent testing and validation have found that the Geostationary Operational Environmental Satellite (GOES) microburst products are effective in the assessment and short-term forecasting of downburst potential and associated wind gust magnitude. Two products, the GOES sounder Microburst Windspeed Potential Index (MWPI) and a new two-channel GOES imager brightness temperature difference (BTD) product have demonstrated capability in downburst potential assessment (Pryor 2009; Pryor 2010). The GOES sounder MWPI algorithm is a predictive linear model developed in the manner exemplified in Caracena and Flueck (1988):

$$MWPI \equiv \{(CAPE/100)\} + \{\Gamma + (T-T_d)_{850} - (T-T_d)_{670}\}$$

where $\Gamma$ is the lapse rate in degrees Celsius (C) per kilometer from the 850 to 670 mb level, and the quantity $(T–T_d)$ is the dewpoint depression (C). In addition, it has been found recently that the BTD between GOES infrared channel 3 (water vapor, 6.5μm) and channel 4 (thermal infrared, 11μm) can highlight regions where severe outflow wind generation (i.e. downbursts, microbursts) is likely due to the channeling of dry mid-tropospheric air into the precipitation core of a deep, moist convective storm. Rabin et.al. (2010) noted that observations have shown that BTD > 0 can occur when water vapor exists above cloud tops in a stably stratified lower stratosphere and thus, BTD > 0 has been used a measure for intensity of overshooting convection. A new feature presented in this paper readily apparent in BTD imagery is a "dry-air notch" that signifies the channeling of dry air into the rear flank of a convective storm.

---

*Corresponding author address*: Kenneth L. Pryor, NOAA/NESDIS/STAR, 5200 Auth Rd., Camp Springs, MD 20746-4304; e-mail: Ken.Pryor@noaa.gov.

## 2. METHODOLOGY

### 2.1 *GOES Imager Microburst Product*

The objective of this validation effort was to qualitatively assess the performance of the GOES imager channel 3-4 BTD in the indication of downburst activity by applying pattern recognition techniques. The image data consisted of derived brightness temperatures from infrared bands 3 and 4, obtained from the Comprehensive Large Array-data Stewardship System (CLASS, http://www.class.ncdc.noaa.gov/). Microburst algorithm output was visualized by McIDAS-V software (version 1.0beta5) (Available online at http://www.ssec.wisc.edu/mcidas/software/v/). A contrast stretch and built-in color enhancement were applied to the output images to highlight patterns of interest including overshooting tops and dry-air notches. The dry-air notch identified in the BTD image is similar in concept to the rear-inflow notch (RIN) as documented in Przybylinski (1995). Visualizing algorithm output in McIDAS-V allowed for cursor interrogation of output brightness temperature difference and more precise recording of BTD values associated with observed downburst events.

For each downburst event, product images were compared to locations of marine transportation accidents, radar reflectivity imagery, and surface observations of convective wind gusts as provided by the Oklahoma and West Texas Mesonets, and WeatherFlow. Next Generation Radar (NEXRAD) base reflectivity imagery acquired from National Climatic Data Center (NCDC) was utilized to verify that observed wind gusts were associated with downbursts and not associated with other types of convective wind phenomena (i.e. gust fronts). Another application of the NEXRAD imagery was to infer microscale physical properties of downburst-producing convective storms. Particular radar reflectivity signatures, such as the bow echo and rear-inflow notch (RIN)(Przybylinski 1995), were effective indicators of the occurrence of downbursts.

## 2.2 GOES Sounder Microburst Product

In a similar manner to the imager product, this validation effort assessed and intercompared the performance of the GOES sounder-derived microburst products by employing classical statistical analysis of real-time data. Accordingly, this effort entailed a study of downburst events over two different microburst environments: the southern Great Plains and the Chesapeake Bay regions. Again, this validation was executed in a manner that emulated historic field projects such as the 1982 Joint Airport Weather Studies (JAWS) (Wakimoto 1985) and the 1986 Microburst and Severe Thunderstorm (MIST) project (Atkins and Wakimoto 1991). Data from the GOES MWPI product was collected over Oklahoma and western Texas for downburst events that occurred between 1 June 2007 and 31 August 2010 and over the Chesapeake Bay region of Maryland and Virginia for events that occurred between 1 June and 31 August 2010. Microburst index values were then validated against surface observations of convective wind gusts as recorded by Oklahoma and West Texas Mesonet stations, and WeatherFlow stations over the Chesapeake Bay region. Wakimoto (1985) and Atkins and Wakimoto (1991) discussed the effectiveness of using mesonet surface observations and radar reflectivity data in the verification of the occurrence of downbursts. Well-defined peaks in wind speed as well as significant temperature decreases (Wakimoto 1985; Atkins and Wakimoto 1991) were effective indicators of high-reflectivity downburst occurrence. As illustrated in the flowchart in Figure 11 of Pryor (2008), images were generated in McIDAS by a program that reads and processes GOES sounder data, calculates and collates microburst risk values, and overlays risk values on GOES imagery. Output images were then archived on the Center for Satellite Applications and Research (STAR) web server.

Since surface data quality is paramount in an effective validation program, Oklahoma, western Texas, and the Chesapeake Bay region were chosen as study regions due to the wealth of high quality surface observation data provided by the Oklahoma and West Texas Mesonets (Brock et al. 1995; Schroeder et al. 2005), and WeatherFlow, and relatively homogeneous topography. Pryor (2008) discussed the importance of the dryline (Schafer 1986) in convective storm climatology in the Southern Plains region as well as the selection of the High Plains region for a validation study.

Similar to the validation process for the imager microburst product, for each microburst event, product images were compared to radar reflectivity imagery. Next Generation Radar (NEXRAD) base reflectivity imagery from NCDC was utilized for this purpose.

Covariance between the variables of interest, MWPI and surface downburst wind gust speed, was considered. Algorithm effectiveness was assessed as the correlation between MWPI values and observed surface wind gust velocities. Statistical significance testing was conducted to determine the confidence level of the correlation between observed downburst wind gust magnitude and microburst risk values.

## 3. CASE STUDIES

### 3.1 *Chesapeake Bay Downbursts*

During the afternoon of 3 June 2010, strong convective storms developed over the Appalachian Mountains of Virginia and West Virginia and then merged to form a squall line over the Maryland and Virginia Piedmont. The squall line produced strong downburst winds over the Tidal Potomac River and Chesapeake Bay between 2100 and 2300 UTC 3 June. The GOES MWPI product effectively indicated the potential for strong downbursts about one hour prior to each event. Surface wind observation data from [WeatherFlow DataScope](#) was instrumental in verifying downburst intensity during this event.

Figures 1 and 2 show that the 2000 UTC MWPI product image indicated elevated values over the entire Chesapeake Bay region downstream of the squall line developing over the Blue Ridge Mountains. The GOES sounding at Baltimore-Washington International (BWI) Airport characterized a favorable "hybrid" environment over the Chesapeake Bay region with large CAPE and dry air layers in the mid-troposphere, near the 400-mb level, and below the cloud base height near the 800-mb level. Thus, precipitation loading and evaporational cooling, both in the storm precipitation core and in the subcloud layer, would be major forcing factors for intense convective downdrafts.

Figure 2 shows the squall line moving eastward through the Maryland and Virginia piedmont and upper Tidal Potomac River between 2000 and 2130 UTC during the time of maximum surface heating when the lower troposphere was most unstable. Figure 1 does show an overall increase in index values over the Chesapeake Bay region between 2000 and 2100 UTC as the lower troposphere was continuing to destabilize. At the same time, as shown in Figure 2, well-defined notches on the rear flank of the squall line signified the channeling of mid-tropospheric dry air into the storm precipitation cores, providing the initial energy for strong convective downdrafts. Pryor (2010) details the physical significance of the dry-air notch in the generation of convective downbursts.

Figure 4 shows composite radar reflectivity images at 2145 UTC (top) and 2220 UTC (bottom), with overlying surface wind observations courtesy of [WeatherFlow DataScope](), Peak wind gusts occurred as the highest radar reflectivity was nearly overhead each observing station. Near 2145 UTC, Potomac Light 33 recorded a sustained wind of 31 knots with a gust to 39 knots followed by a sustained wind of 27 knots with a gust to 41 knots near 2220 UTC at Kent Island. Note the orientation of a dry-air notch toward Potomac Light 33 station at 2132 UTC, just prior to the observance of a downburst. The 2100 UTC MWPI product displayed index values of 20 to 24 in proximity to the locations of downburst observation. MWPI values of 20 to 24 correspond to wind gust potential of 39 to 41 knots. During this event, the MWPI product demonstrated effectiveness with a close correspondence between index values and wind gust magnitude. The BTD product images in Figure 2 emphasized the role of mid-tropospheric dry air and its interaction with the convective storm line in the generation of intense downdrafts.

### 3.2 *West Texas Downbursts*

During the afternoon of 22 June 2010, an upper-level disturbance interacted with a dryline over eastern New Mexico and triggered strong convective storms that tracked eastward into the western Texas Panhandle region. Convective storms produced scattered strong downbursts over western Texas during the evening hours. The 2200 and 2300 UTC GOES MWPI product images effectively indicated the potential for strong downbursts two to three hours prior to each event. Surface wind observation data from [West Texas Mesonet]() was instrumental in verifying downburst intensity during this event.

The MWPI product performed especially well during the evening of 22 June. Figure 5 shows that the 2200 and 2300 UTC MWPI product images indicated elevated values near the Texas-New Mexico border, west of longitude 102°W, and a local maximum near Muleshoe, TX at 2300 UTC (orange to red shading) with index values of 40 to 45. A line of convective storms developed over eastern New Mexico and then tracked into western Texas after 0000 UTC. The corresponding GOES sounding profile in Figure 5 displays a favorable classic "inverted V" profile that prevailed over western Texas. A deep dry adiabatic lapse rate (DALR) layer below the 670-mb level and a dry subcloud layer fostered intense downdraft development due to evaporational cooling and resultant negative buoyancy generation. Upon impact of the intense downdrafts on the surface, strong downburst winds were generated. The MWPI product was particularly effective as index values of 40 to 45 corresponded to measured downburst wind gusts of 46 knots at Hereford at 0110 UTC and 43 knots at Muleshoe at 0120 UTC 23 June. Based on previously established statistical relationship, MWPI values of 40 to 45 correspond to wind gust potential of 46 to 48 knots.

As the squall line tracked eastward over the Texas-New Mexico border, particularly well defined dry-air notches became apparent over eastern New Mexico on the western flank of the line. The dry-air notches were oriented toward the east and toward intense cells that were producing strong downburst winds over Hereford and Muleshoe. As in the previous case, these dry-air notches, in line with spearhead echoes (Fujita and Byers 1977) as identified in radar reflectivity imagery, signified the channeling of dry mid-tropospheric air into heavy precipitation cores within the squall line. The interaction of this dry air with the heavy precipitation resulted in the generation of strong negative buoyancy and subsequent intense downdrafts. Also noteworthy is the correlation between inflow region BTD values and magnitude of downburst wind gusts as illustrated in Figure 7. The BTD in this case likely indicated the relative dryness of the inflow mid-tropospheric air that enhanced downdraft intensity as the squall line tracked through western Texas.

### 3.3 *Oklahoma Downbursts*

On 20 and 21 August 2010, a weak cold front drifted southward over Oklahoma and served as a trigger for strong single cell and multicell convective storms that developed during the evening hours. GOES-derived MWPI product images on both 20 and 21 August effectively indicated the potential for strong downbursts with wind gusts of 43 to 49 knots over north-central and south-central Oklahoma, respectively. A strong downburst wind gust of 47 knots was recorded at Lahoma mesonet station at 0145 UTC 21 August, followed by a stronger downburst wind gust of 56 knots that was recorded at Ringling mesonet station at 2345 UTC. The channeling of mid-tropospheric dry air into the heavy precipitation core of the storms was most likely a significant factor in downburst magnitude, especially at Ringling.

Figure 8 shows that the MWPI product images indicated elevated wind gust potential (43 to 49 knots, blue to orange-colored markers) over central Oklahoma on 20 August and over southern Oklahoma on 21 August, in the vicinity of developing convective storm activity along the cold front boundary. Note higher index values (45 to 48, yellow marker) in proximity to the occurrence of the stronger downburst wind gust at Ringling (56 knots) than that indicated near Lahoma (green marker). Although cloud cover associated with the frontal boundary was more widespread during the afternoon and evening of 21 August, resulting in a more sporadic areal coverage of sounding retrievals over southern Oklahoma, a local maximum in MWPI values (47, yellow marker) was apparent on the inflow (southwest) side of the downburst-producing storm over Ringling.

As shown in Figure 9, dry-air notches readily apparent in the GOES BTD images near the time of downburst occurrence at 0145 and 2345 UTC, respectively, were associated with the observed downburst winds. Note that larger BTD values, indicated by the dark blue shading in the mid-tropospheric inflow region, were associated with the stronger downburst that was observed at Ringling. The slight displacement between the satellite-indicated dry-air notches and associated RINs as detected by NEXRAD is most likely the result of error introduced by the large distance between the storms and the GOES-13 subpoint at latitude 0°N, longitude 75°W. As expected from geometric calculations, the displacement error over Oklahoma typically ranges from seven to eight nautical miles. The presence of a dry-air notch in satellite imagery signified the channeling of dry mid-tropospheric air into the rear flank of the convective storms and the interaction with the heavy precipitation cores that provided energy for strong downdrafts and the resulting downburst winds.

## 4. STATISTICAL ANALYSIS AND DISCUSSION

Validation results for the 2007 to 2010 convective seasons have been completed for the MWPI and imager microburst products. GOES sounder-derived MWPI values have been compared to mesonet observations of downburst winds over Oklahoma and Texas for 208 events between June 2007 and September 2010. The correlation between MWPI values and measured wind gusts was determined to be .62 and was found to be statistically significant near the 100% confidence level, indicating that the correlation represents a physical relationship between MWPI values and downburst magnitude and is not an artifact of the sampling process. Figure 10 shows a scatterplot of MWPI values versus observed downburst wind gust speed as recorded by mesonet stations in Oklahoma and Texas. The MWPI scatterplot identifies two clusters of values: MWPI values less than 50 that correspond to observed wind gusts of 35 to 50 knots, and MWPI values greater than 50 that correspond to observed wind gusts of greater than 50 knots. The scatterplot illustrates the effectiveness of the MWPI product in distinguishing between severe and non-severe convective wind gust potential. In addition, for the Chesapeake Bay and Tidal Potomac River region, correlation computed between MWPI values and measured wind gusts was .58. This correlation, based on seven downburst events, was found to be statistically significant with a confidence level near 100%.

Further statistical analysis of a dataset built by comparing wind gust speeds recorded by Oklahoma Mesonet stations to adjacent MWPI values for 35 downburst events has yielded some favorable results, as displayed in Figures 11 and 12. Correlation was computed between key parameters in the downburst process, including temperature lapse rate (LR) and dewpoint depression difference (DDD) between two levels (670mb/850mb), CAPE, and radar reflectivity. The first important finding is a statistically significant negative correlation ($r=-.34$) between lapse rate and radar reflectivity. Similar to the findings of Srivastava (1985), for lapse rates greater than 8 K/km, downburst occurrence is nearly independent of radar reflectivity. For lapse rates less than 8 K/km, downburst occurrence was associated with high reflectivity (> 50 dBZ) storms. The majority of downbursts occurred in sub-cloud environments with lapse rates greater than 8.5 K/km. Adding the dewpoint depression difference to lapse rate yielded an even greater negative correlation (r=-.42) when compared to radar reflectivity. Finally, comparing the sum of LR and DDD (the former hybrid microburst index (HMI)) to CAPE resulted in the strongest negative correlation (r=-.82), with a confidence level above 99%. This emphasizes the complementary nature of the HMI and CAPE in generating a robust and physically meaningful MWPI value. This result also shows that CAPE can serve as an adequate proxy variable for precipitation loading (expressed as radar reflectivity) in the MWPI. The strong negative correlation, or negative functional relationship, between HMI and CAPE terms in the MWPI algorithm indicates that the MWPI should be effective in capturing both negative buoyancy generation and precipitation loading as downburst forcing mechanisms.

## 5. SUMMARY AND CONCLUSIONS

As documented in Pryor (2009, 2010), and proven by statistical analysis, the GOES sounder MWPI product has demonstrated capability in the assessment of wind gust potential over the southern Great Plains and Chesapeake Bay regions. Case studies and statistical analysis for downburst events that occurred during the 2007 to 2010 convective seasons demonstrated the effectiveness of the GOES microburst products. However, as noted by Caracena and Flueck (1988), the majority of microburst days during JAWS were characterized by environments intermediate between the dry and wet extremes (i.e. hybrid). As noted in Pryor (2009), the MWPI product is especially useful in the inference of the presence of intermediate or "hybrid" microburst environments, especially over the Great Plains region. Further validation over the Chesapeake Bay region should strengthen the functional relationship between MWPI values and downburst wind gust magnitude.

The dry-air notch identified in all three case studies presented above most likely represents drier (lower relative humidity) air that is channeled into the rear of convective storms and interacts with the precipitation cores, subsequently providing the energy for intense downdrafts and resulting downburst winds. Comparison of BTD product imagery to corresponding radar imagery revealed a correlation between the dry-air notch and the RIN. Entrainment of drier mid-tropospheric air into the precipitation core of the convective storm typically results in evaporation of precipitation, the subsequent cooling and generation of negative buoyancy (sinking air), and resultant acceleration of a downdraft. When the intense localized downdraft reaches the surface, air flows outward as a downburst. Ellrod (1989) noted the importance of low mid-tropospheric (500 mb) relative humidity air in the generation of the severe Dallas-Fort Worth, Texas microburst in August 1985. Thus, the channel 3-4 BTD product can serve as an effective supplement to the GOES sounder MWPI product. Further validation of the imager microburst product and quantitative statistical analysis to assess product performance will serve as future work in the development and evolution of the GOES microburst products.

**Acknowledgements**

The author thanks the Oklahoma Mesonet, the West Texas Mesonet, and Jay Titlow (WeatherFlow) for the surface weather observation data used in this research effort. The author thanks Michael Grossberg and Paul Alabi (NOAA/CREST, CCNY) for their implementation of the MWPI program into the Graphyte Toolkit and their assistance in generating the MWPI product images. The author also thanks Jaime Daniels (NESDIS) for providing GOES sounding retrievals displayed in this paper.

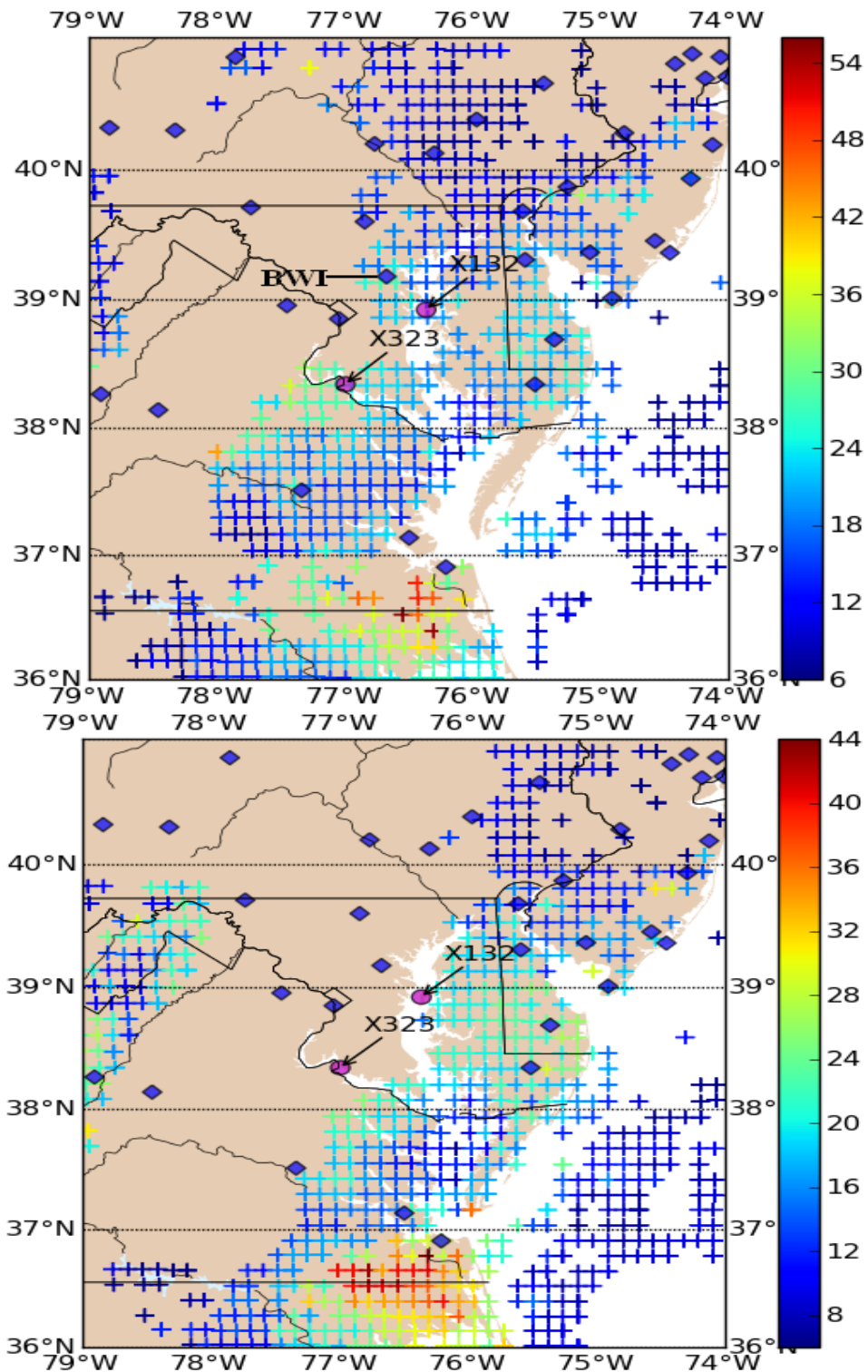

Figure 1. GOES sounder MWPI products at 2000 UTC (top) and 2100 UTC (bottom) 3 June 2010 visualized by the Graphyte toolkit.

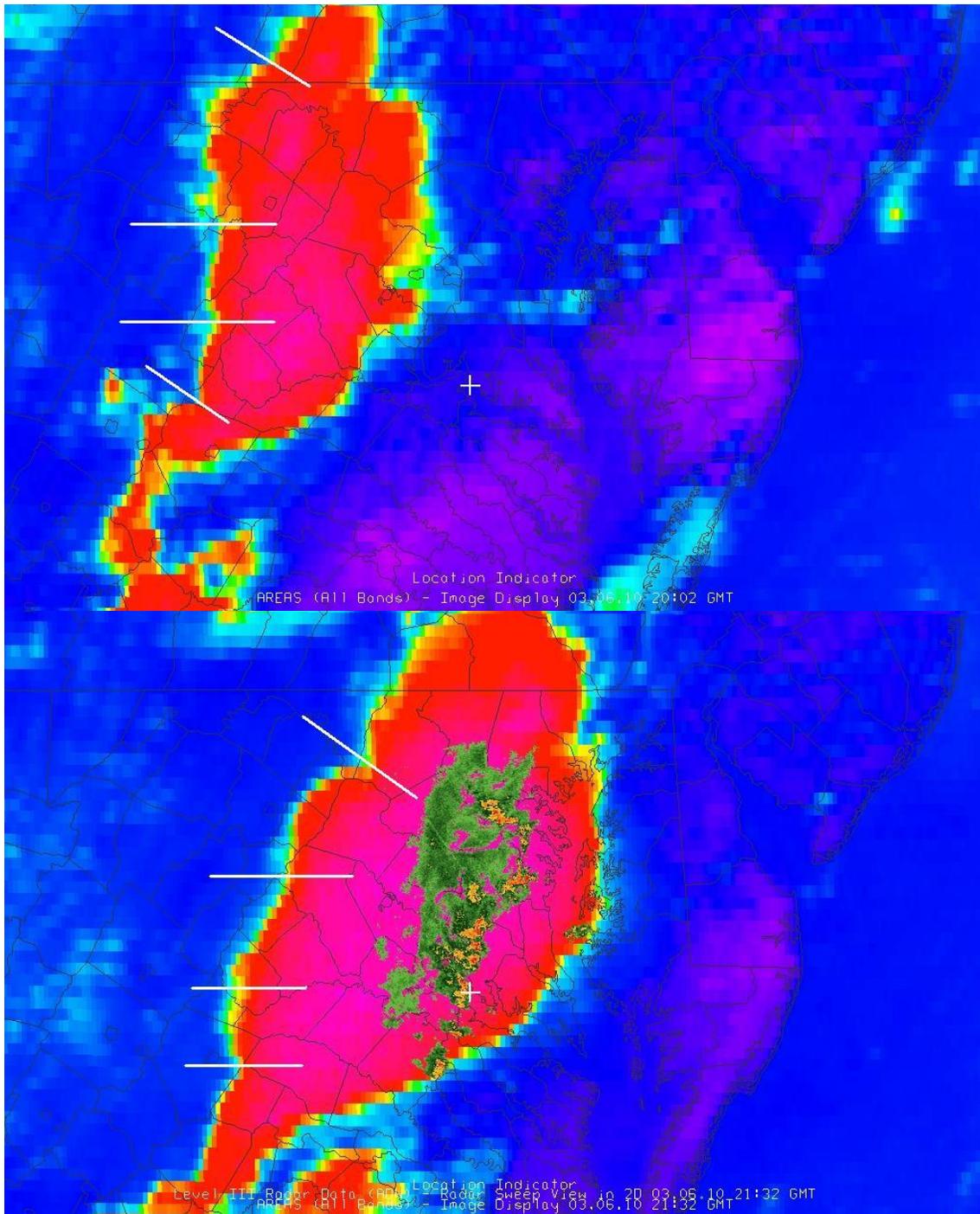
Figure 2. GOES imager channel 3-channel 4 brightness temperature difference (BTD) image products at 2002 UTC (top) and 2132 UTC (bottom) 3 June 2010 with overlying radar reflectivity from Andrews Air Force Base TDWR. White cross marks location of Potomac Light 33 observing station. White lines represent dry-air channels into the rear flank of the squall line.

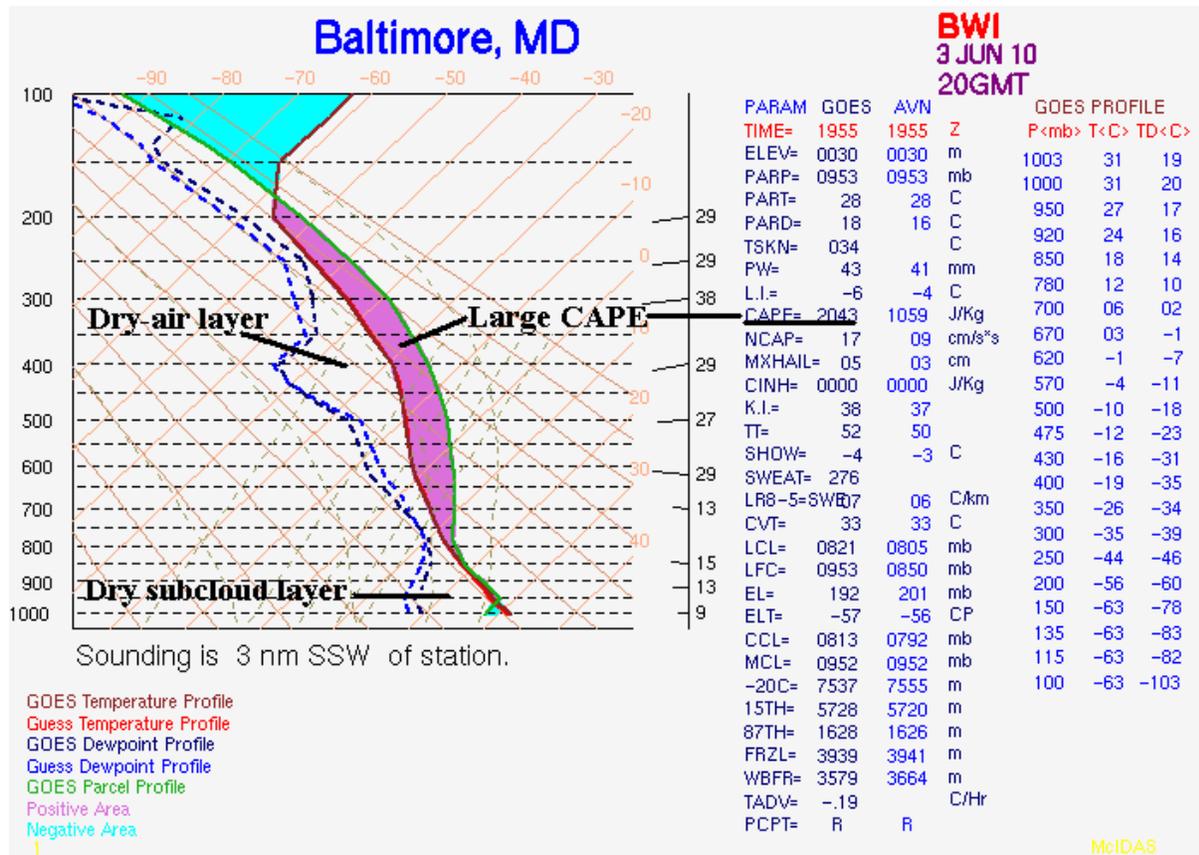

Figure 3. Geostationary Operational Environmental Satellite (GOES) sounding profile from Baltimore, Maryland (BWI) at 2000 UTC 3 June 2010.

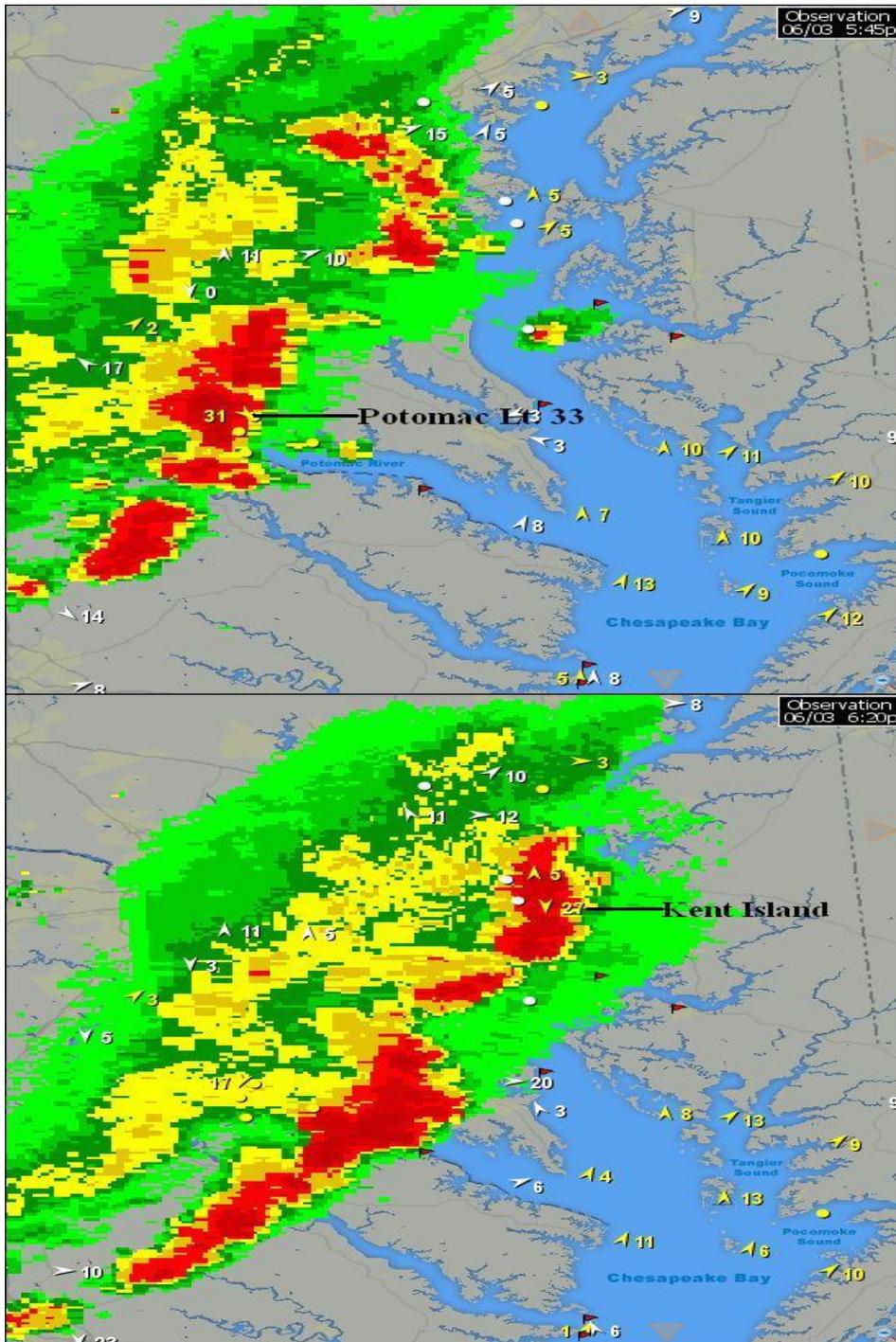
Figure 4. Composite radar reflectivity images at 2145 UTC (top) and 2220 UTC (bottom), with overlying surface wind observations, at the time of downburst occurrence at Potomac Light 33 and Kent Island stations, respectively. Images are courtesy of WeatherFlow DataScope.

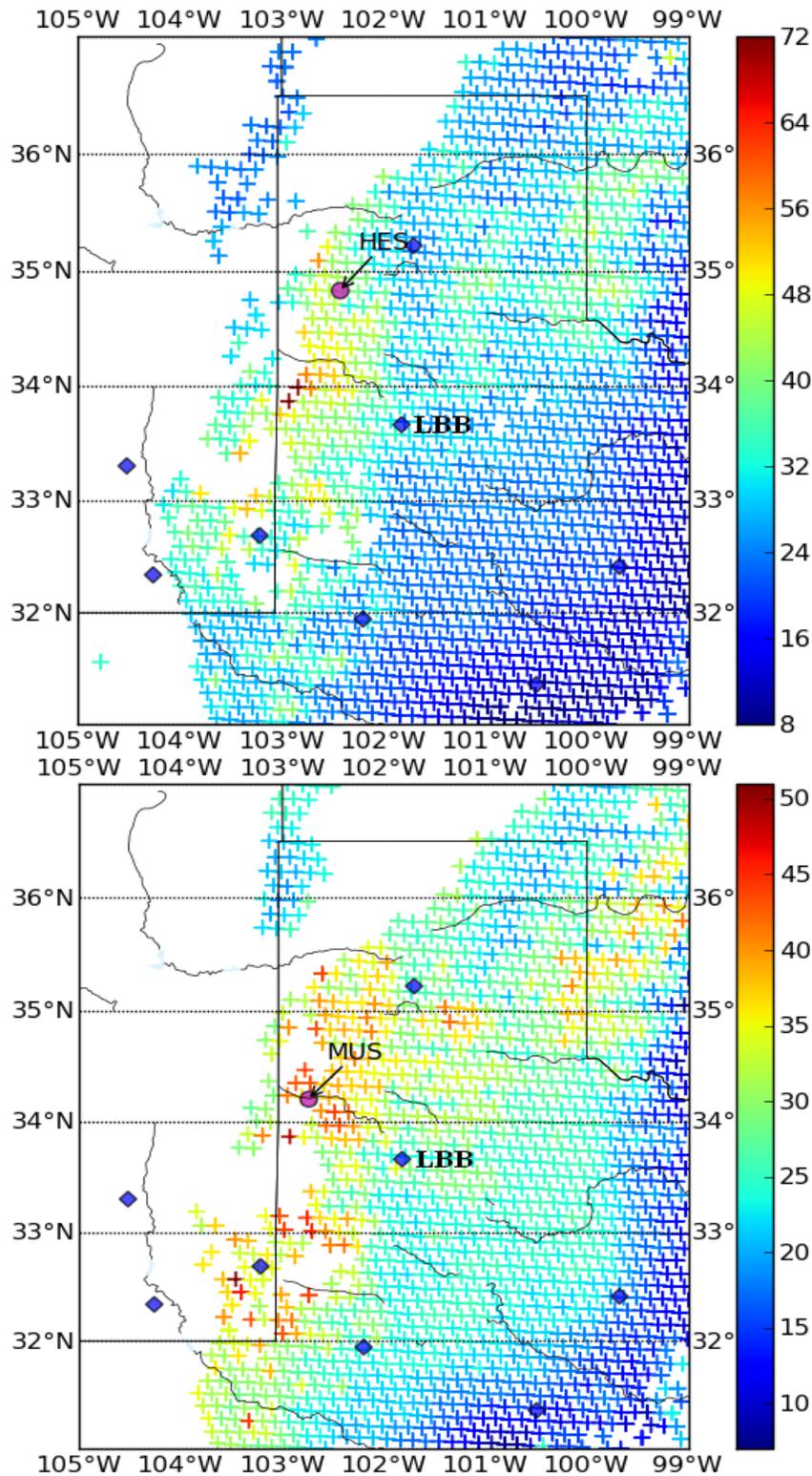

Figure 5. Geostationary Operational Environmental Satellite (GOES) MWPI product images at 2200 UTC (top) and 2300 UTC (bottom) 22 June. Locations of West Texas Mesonet observations of downburst winds (HES, MUS) are plotted on the image.

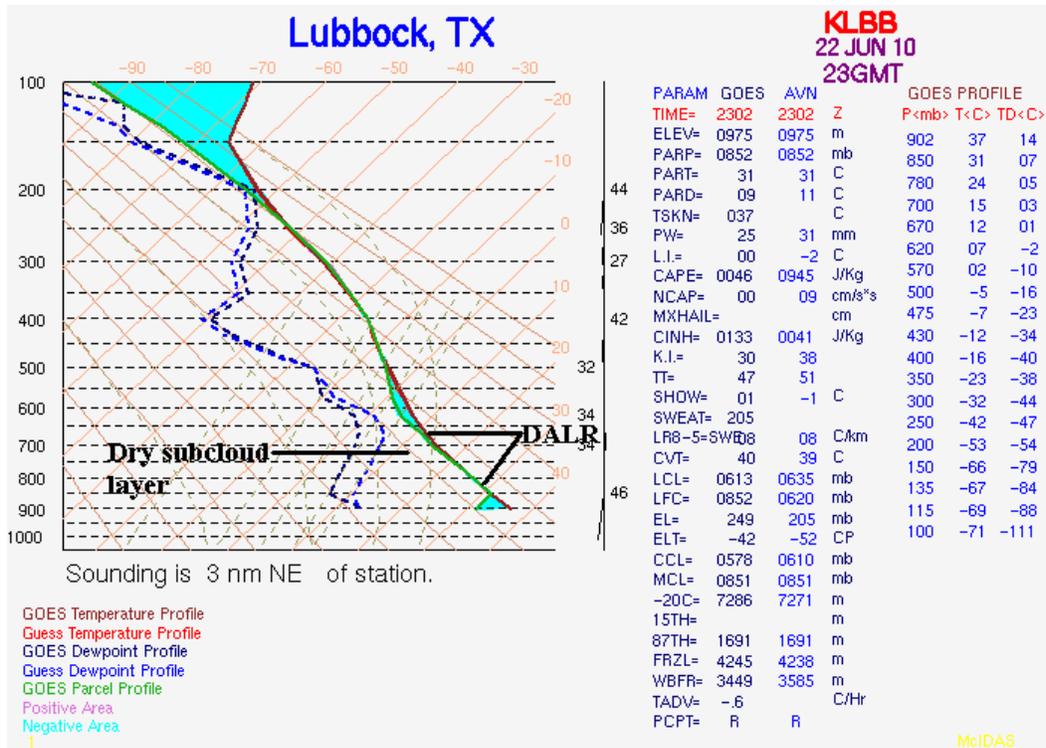

Figure 6. Geostationary Operational Environmental Satellite (GOES) sounding profile from Lubbock, Texas at 2300 UTC 22 June 2010 (bottom).

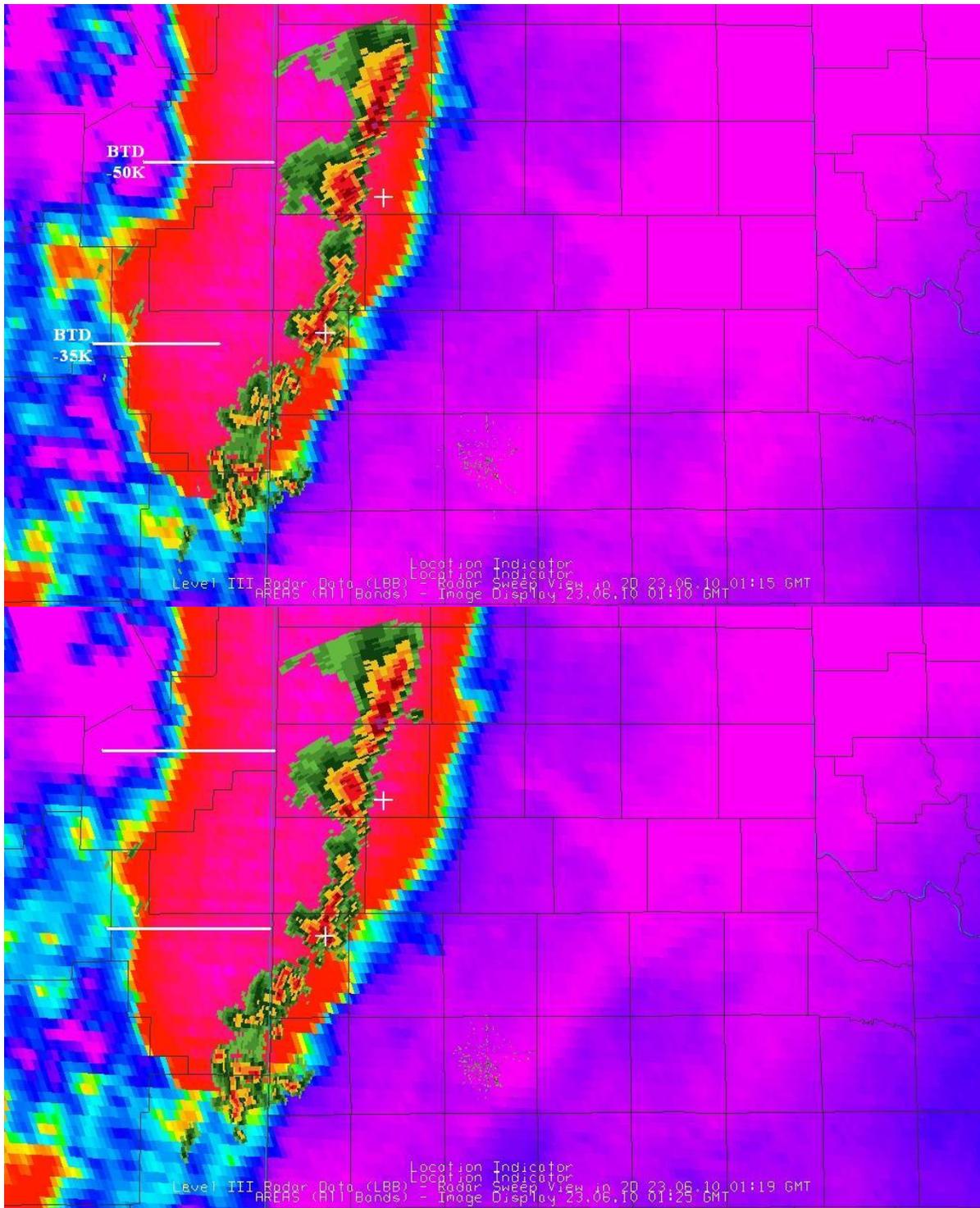

Figure 7. GOES-13 channel 3 (WV)-channel 4 (IR) brightness temperature difference (BTD) product images at 0115 UTC (top) and 0119 UTC (bottom) 23 July 2010 with overlying radar reflectivity from Lubbock, Texas NEXRAD. White lines represent the dry-air notches pointing to Hereford (northern white cross) and Muleshoe (southern white cross) West Texas Mesonet stations.

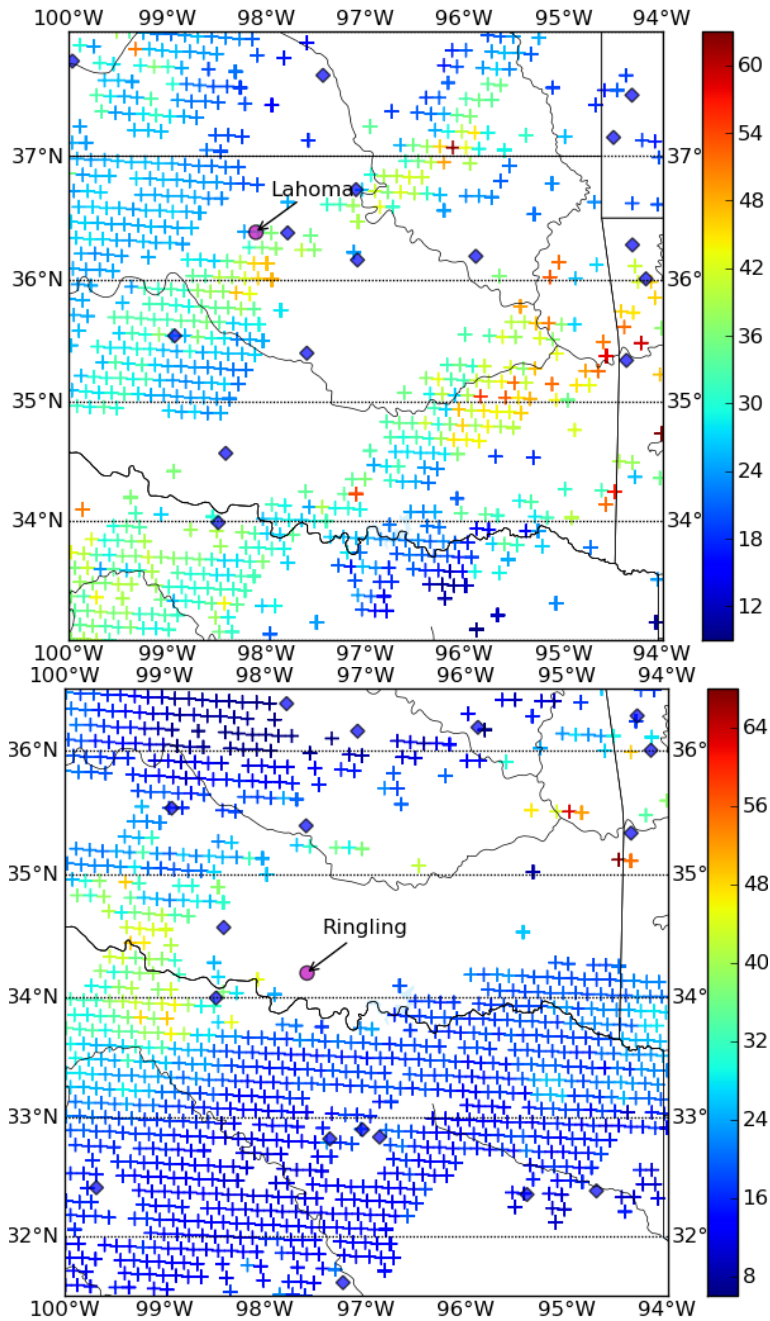

Figure 8. Geostationary Operational Environmental Satellite (GOES) MWPI product images, visualized by the Graphyte Toolkit, at 2300 UTC 20 August 2010 (top) and 2200 UTC 21 August (bottom). The locations of the observation of downburst winds are plotted on the images.

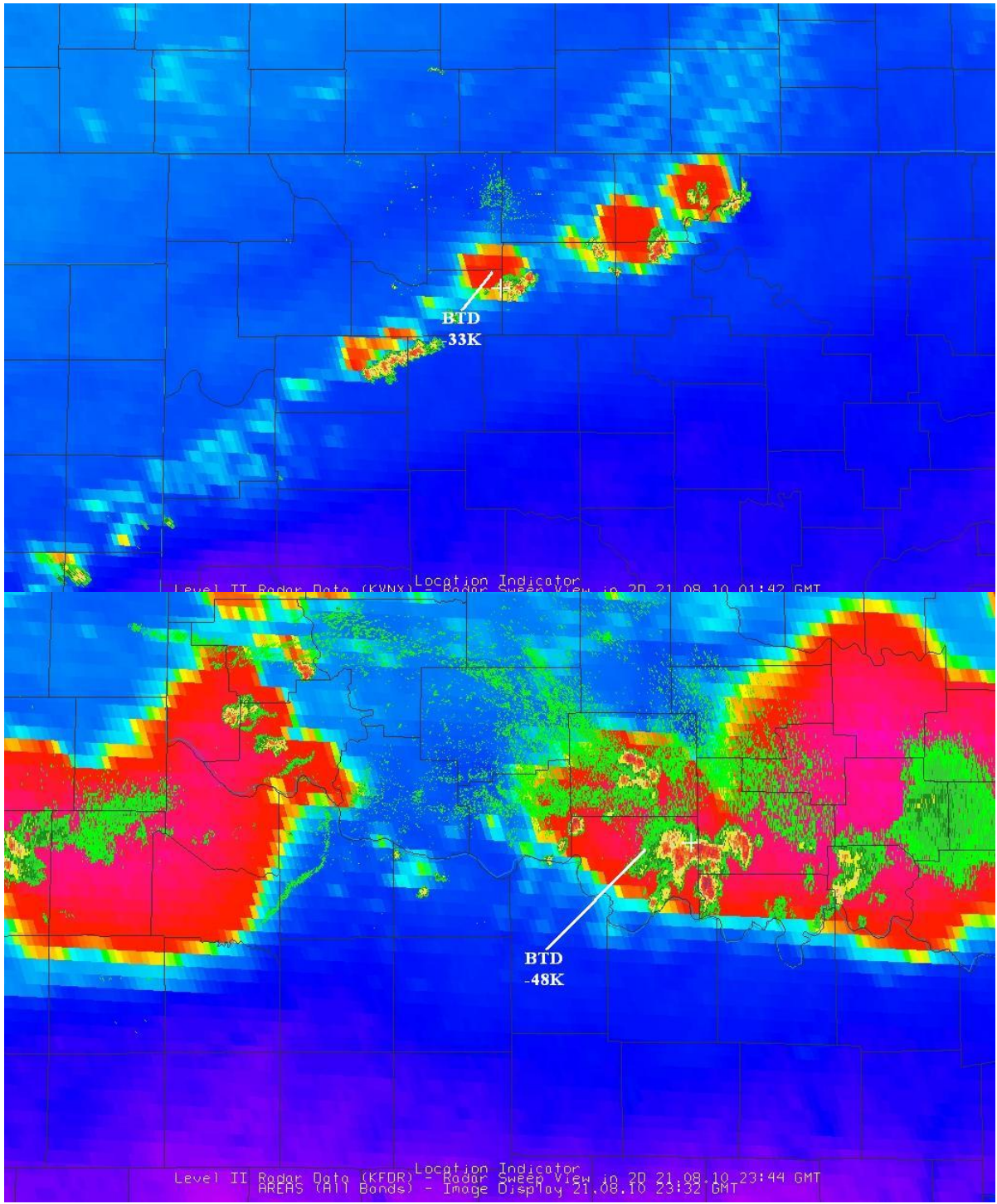

Figure 9. GOES channel 3-4 brightness temperature difference (BTD) products at 0132 (top) and 2345 UTC (bottom) 21 August 2010 with overlying radar reflectivity. White lines represent dry-air notches pointing to Lahoma and Ringling Oklahoma Mesonet stations. White crosses mark the locations of downburst observation at Lahoma (top) and Ringling (bottom), respectively.

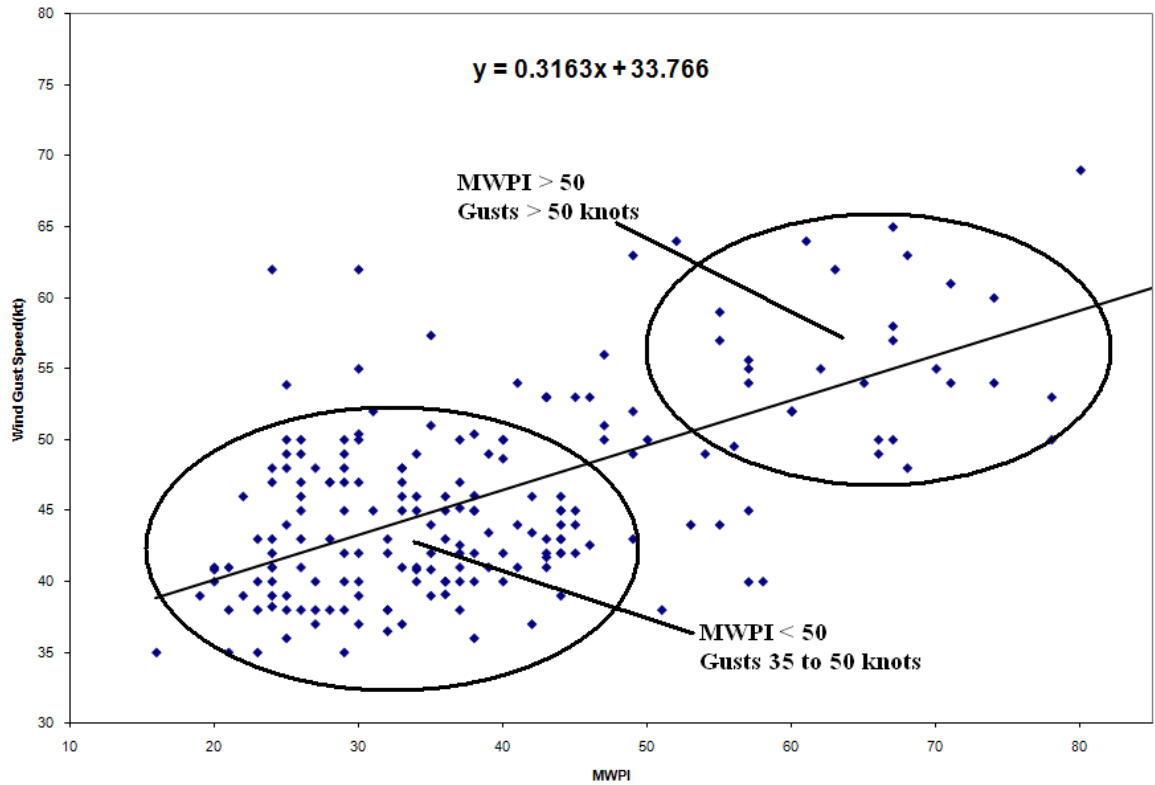

Figure 10. Statistical analysis of validation data over the Oklahoma and western Texas domain between June and September 2007 through 2010: Scatterplot of MWPI values vs. measured convective wind gusts for 208 downburst events.

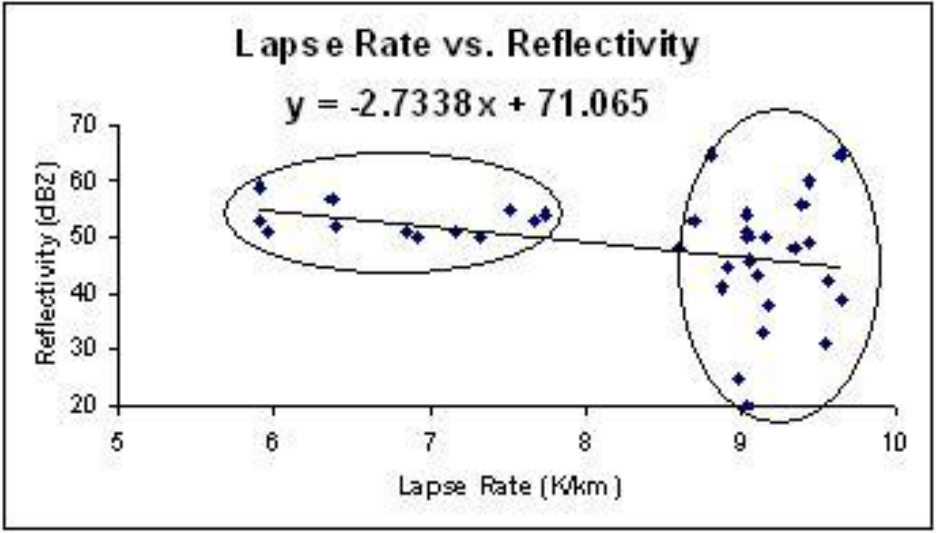
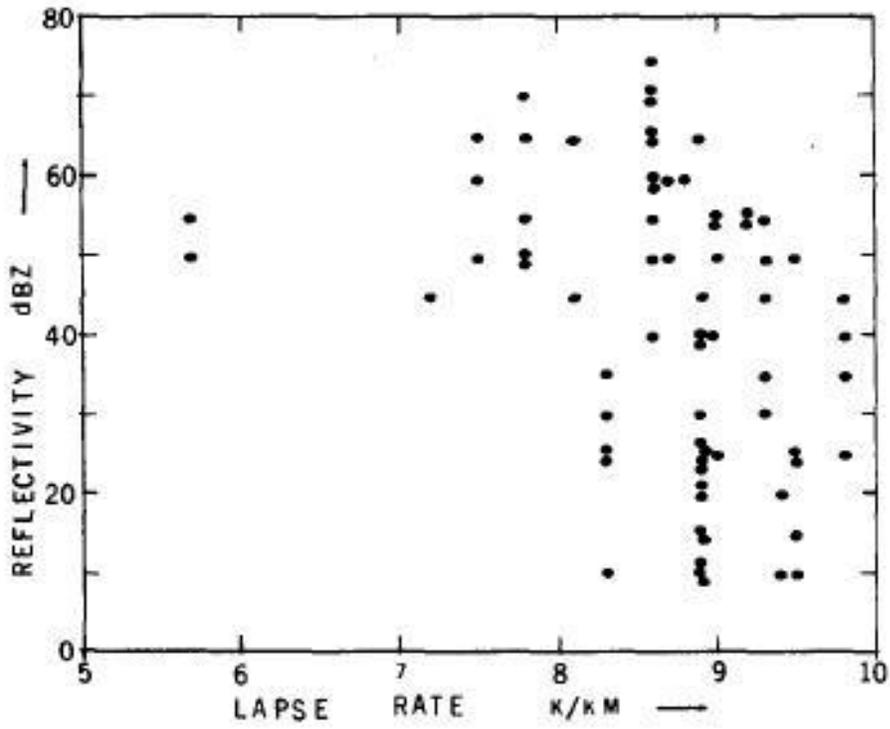

Figure 11. Scatterplot of lapse rate versus radar reflectivity for 35 downburst events over Oklahoma during the 2009 convective season (top) compared to scatterplot for 186 microburst events during the 1982 JAWS project (courtesy Srivastava 1985).

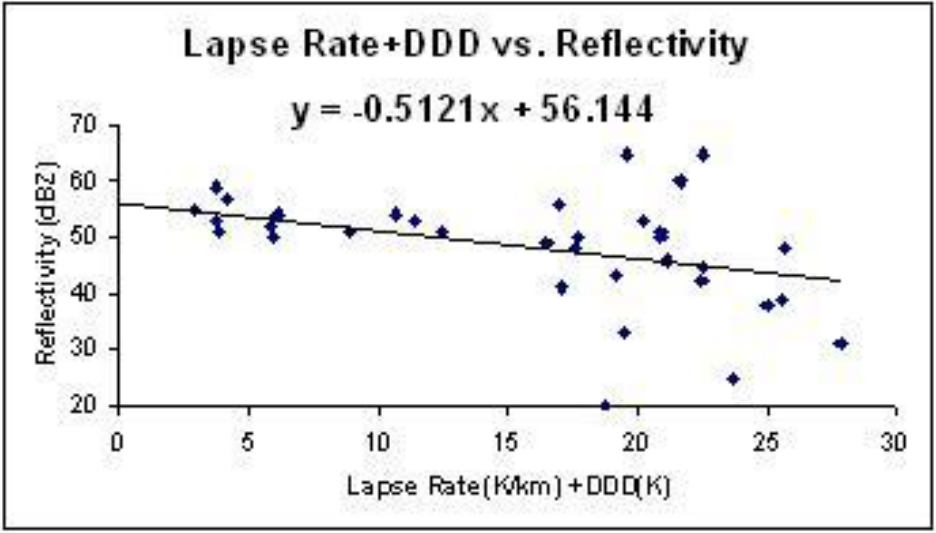

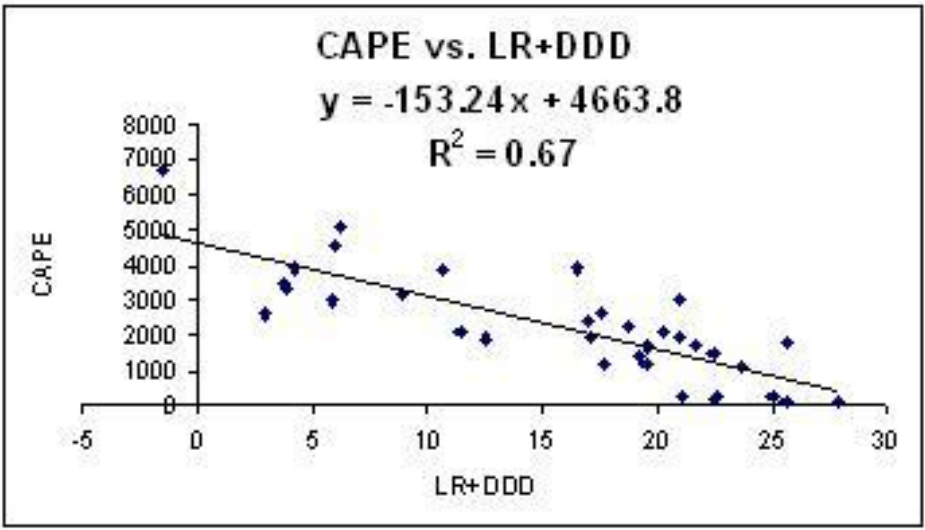

Figure 12. Scatterplot of the sum of lapse rate and DDD versus radar reflectivity (top) compared to scatterplot of the sum of lapse rate and DDD versus CAPE (bottom) for 35 downburst events over Oklahoma during the 2009 convective season.